# Opportunities and Challenges in Harnessing Digital Technology for Effective Teaching and Learning

Zhongzhou Chen [1,*] and Chandralekha Singh [2]

[1] Department of Physics, University of Central Florida, Orlando, FL 32816, USA
[2] Department of Physics and Astronomy, University of Pittsburgh, Pittsburgh, PA 15260, USA; clsingh@pitt.edu
* Correspondence: zhongzhou.chen@ucf.edu

**Abstract:** Most of today's educators are in no shortage of digital and online learning technologies available at their fingertips, ranging from Learning Management Systems such as Canvas, Blackboard, or Moodle, online meeting tools, online homework, and tutoring systems, exam proctoring platforms, computer simulations, and even virtual reality/augmented reality technologies. Furthermore, with the rapid development and wide availability of generative artificial intelligence (GenAI) services such as ChatGPT, we are just at the beginning of harnessing their potential to transform higher education. Yet, facing the large number of available options provided by cutting-edge technology, an imminent question on the mind of most educators is the following: how should I choose the technologies and integrate them into my teaching process so that they would best support student learning? We contemplate over these types of important and timely questions and share our reflections on evidence-based approaches to harnessing digital learning tools using a Self-regulated Engaged Learning Framework we have employed in our research in physics education that can be valuable for educators in other disciplines.

## 1. Introduction

The development of educational theory and practice has long emphasized the importance of considering multiple dimensions in designing effective learning environments. From Dewey's emphasis on experiential learning [1] to Vygotsky's sociocultural theory [2] and Bandura's work on self-efficacy [3], scholars have recognized that successful learning requires attention to both internal cognitive processes and external environmental factors. In 2020 and 2021, when almost every instructor was forced to adopt digital learning during the COVID-19 pandemic [4–16], a significant body of studies has shown that students' learning experiences varied drastically in online settings depending on the selection and implementation of digital learning technologies. In addition, many Learning Management Systems [17], online homework and tutoring systems, online meeting tools (e.g., Zoom), exam proctoring platforms, computer simulations [18–22], and virtual reality/augmented reality technologies [23–28] provide instructors with lots of freedom to design the learning experience for their students [29–33]. Moreover, in the last two years, the emerging potential of generative artificial intelligence (GenAI), e.g., large language models (LLMs) such as ChatGPT, in enhancing and personalizing learning is beginning to be recognized as an exciting frontier for transforming education for students at all levels in all disciplines worldwide [34–47]. While these recent advances in digital learning technology options present opportunities to provide effective education to a variety of learners from diverse backgrounds [48], the potential of technology can never be realized without pedagogical considerations of the in-person, remote, or hybrid [49] courses. Therefore, in this article, we, two researchers in physics education, share our reflections on harnessing digital learning technologies pedagogically to ensure that students from diverse backgrounds can benefit from the wide range of digital tools that can be used to enhance learning across different disciplines.

Before diving into the details of the framework for the selection and implementation of digital learning technology, we will first clarify what we mean by "digital learning" in this article, and what distinguishes it from all of its predecessors such as DVDs, video tapes, and TV [50–55]. We use "digital learning" to refer to any computer-based learning system that allows and requires the user to interact with it as an integral part of their learning experience [56–62]. For example, an online homework platform is a digital learning system because submitting an answer to a problem is an interaction that is essential to the learning process. In contrast, we do not consider a standalone static webpage or a lecture video to be a digital learning tool since the learners do not interact with them beyond passively reading or watching. Rather than "digital learning tools", those types of resources might be simply referred to as "digital broadcasting" tools that are similar in nature to video tapes and TV. On the other hand, if the webpage or video is embedded within a system that also includes assessment tasks or online discussion that students must engage with, then they could be considered as part of a digital learning system. The self-paced and interactive nature of many digital learning tools makes it possible for today's higher education students with diverse backgrounds, learning needs, and levels of prior preparation to engage with learning activities at different times, locations, and paces, and possibly with customized content and personalized feedback. These aspects of digital learning tools are central for supporting equity in learning. At the same time, these self-paced digital learning tools also require a certain level of self-regulation from students for them to be used effectively and be conducive to learning [63].

For instructors, evidence-based self-paced digital learning tools can provide a much higher level of flexibility in the design of learning experiences for students with diverse prior preparations. This can be achieved by adjusting a wide range of parameters ranging from simple options such as flexible due dates, number of allowed attempts, and course credit incentives to sophisticated features such as adaptive feedback based upon students' level of mastery, the structure of discussion forums, and game rules in gamified learning environments. On the other hand, this high level of flexibility also places an overwhelming burden on the instructor to find the optimal design and implementation of digital learning experiences that would result in a meaningful and engaging learning experience for students from diverse backgrounds. From both our own extensive prior research experience, as well as experience during the past few years since 2020 throughout the pandemic, even the most advanced digital tools can end up not benefiting or even harming student learning when not properly designed or implemented with student characteristics in mind. For example, long synchronous online lectures delivered via Zoom can be even worse than their in-person counterparts if no evidence-based active-engagement activities are incorporated together with the lectures [64–66].

## 2. Framework for Self-Paced Digital Learning

We first discuss a framework to help educators think systematically about how to select and implement digital learning tools in their courses, supplemented with multiple examples from our research and classroom implementations [67–70]. The framework, called Self-regulated Engaged Learning Framework (SELF) (see Figure 1), synthesizes decades of research on self-regulated learning [67–70], cognitive psychology [2,71–74], and technology-enhanced education [75–77] into a holistic approach for the digital age.

While building on established educational principles, SELF is uniquely focused on addressing the challenges that arise when learning is mediated through digital tools—particularly with the increased demands for self-regulation and the potential for both enhanced engagement and disengagement that technology can create. It is a holistic framework that supports self-regulated learning and consists of four interrelated dimensions that collectively determine how effectively students engage with and learn from learning tools. Each dimension addresses specific aspects of the learning environment while acknowledging their interconnections and suggests that an effective design of learning tools (factor I), their implementation (factor II), student [78] internal characteristics (factor III), and their social and environmental factors (factor IV) collectively determine how effectively students from diverse backgrounds engage with and learn from instructional tools. Thus, while the framework is valuable for self-regulated learning in general, it is applicable to learning using almost all digital learning tools, as the often self-paced and interactive nature of digital learning inevitably requires a certain level of self-regulation from students.

|  | Learning Tool Characteristics | Student Characteristics |
|---|---|---|
| Internal Characteristics | **Factor I.** Learning tool characteristics for self-regulated learning (internal)—pertaining to how effectively the tool focuses on knowledge/skills to be learned | **Factor II.** Student characteristics (internal), e.g., their prior knowledge, skills, identity, and beliefs relevant to engagement and learning |
| External Characteristics | **Factor III.** Learning tool characteristics for self-regulated learning (external)—pertaining to how the tool is implemented and incentivized to optimize student motivation and self-efficacy to engage with it as part of a course | **Factor IV.** Student characteristics (external)—pertaining to student–environment interaction, e.g., role of community/instructor support and other human elements including encouragement for engagement with the tool |

**Figure 1.** Self-regulated Engaged Learning Framework (SELF) (adapted from Ref. [67]).

## 3. Factors That Allow Digital Learning Tools to Promote Self-Regulated Learning

Here, we will illustrate the use of this holistic framework for our own research in physics education in the context of optimizing student engagement with, and learning from, digital learning tools [67–70]. Figure 1 shows that the framework consists of four quadrants, with each quadrant corresponding to one factor:

**Factor I:** the internal characteristics of a digital learning tool pertain to the features of the tool itself, e.g., how effective and adaptive the tool is for knowledge and skills to be learned by students for whom they are designed (whether the tool is evidence-based and includes formative assessment).

**Factor II:** the external characteristics of a digital learning tool pertain to how the tool is implemented as part of a course or a learning environment (e.g., whether the tool is integrated appropriately as part of the instructional design of a course, implemented to account for students' self-efficacy, and incentivized to obtain students' buy-in and engagement).

**Factor III:** the internal characteristics of the students pertain to their relevant prior preparation, knowledge, skills, identity, and beliefs (e.g., self-efficacy [3]) relevant for engagement with and learning from digital tools.

**Factor IV:** the external characteristics of the students pertain to social and environmental factors such as encouragement and community support, e.g., from the course instructor and peers, for engaging with digital learning tools, and managing time to balance the multiple demands of everyday life.

In Figure 1, the four dimensions or quadrants focus on the characteristics of the learning tools and students. The top and bottom rows focus on the internal and external characteristics of the learning tools and the students. In particular, factors I and III involve the internal and external characteristics of digital learning tools (e.g., how they should be evidence-based and have adaptive features that provide opportunities to students with diverse prior knowledge and skills to benefit from them) and how the tools are implemented and incentivized to help students with diverse prior preparations, skills, identities, self-efficacy, and beliefs to engage with them. Factors II and IV involve student characteristics. To use digital learning tools to create a learning environment that supports effective and equitable learning for a diverse group of students, all four quadrants must be considered holistically. In the rest of this paper, we will present several examples from physics of how consideration of all four factors, especially factors III and IV in the lower half that are often ignored by instructors, could lead to improved learning for students with diverse prior preparations.

## 4. Examples of Consideration of SELF Framework for Improving Digital Learning

*4.1. Mastery-Based Online Learning Modules: A Design That Considers Both the Internal and External Characteristics of the Tools*

To design effective digital learning tools, educators often first consider the internal characteristics of the tools and students, i.e., factors I and II in the SELF framework. For example, Schwartz, Bransford, and Sears' [79] preparation for the future learning model emphasizes that for students to engage appropriately with learning tools, there should be elements of both efficiency and innovation embedded in the instructional tools and design. One interpretation of efficiency and innovation in this model is that if the students are asked to engage with learning tools that are too efficient and easy, students will disengage, and learning will not be meaningful; on the other hand, if the learning tools are too innovative and challenging, students will struggle so much while engaging with them that they will get frustrated and give up. Thus, digital learning tools should have an appropriate blending of both efficiency and innovation so that students engage and struggle productively while learning. Ensuring an appropriate balance of efficiency and innovation with a focus on equity requires that digital learning tools be able to provide a productive learning experience for students with diverse prior knowledge and skills (factor II). For example, effective digital learning tools can have formative assessments built into them so that students can receive continuous feedback as they engage with them and evaluate their own learning as they make progress.

We will use our design of mastery-based online learning modules [80,81] for introductory physics as an example of how factors I and II are considered during the design of a digital learning tool, as illustrated in Figure 2.

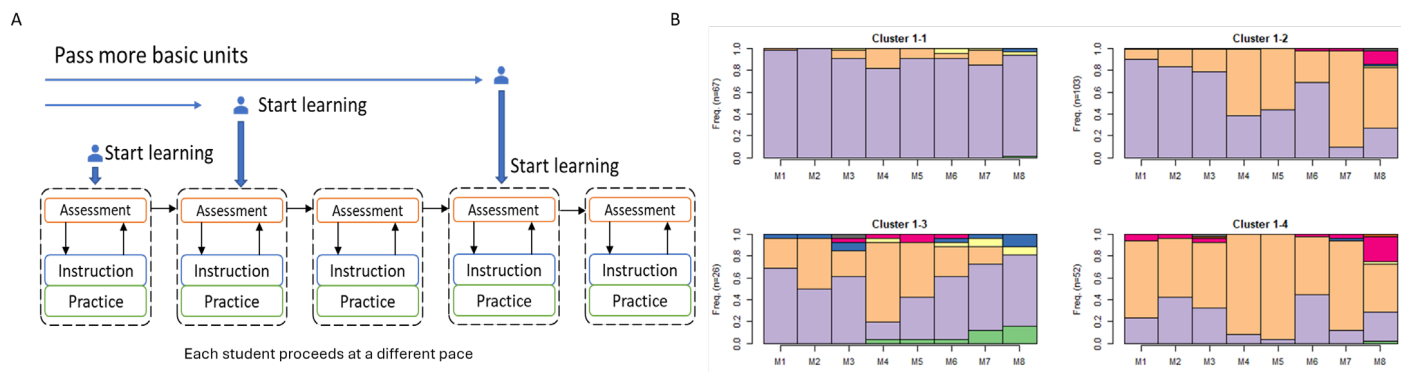

**Figure 2.** Mastery-based online learning modules. (**A**) Schematic illustration of the mastery-based online learning modules design. (**B**) Four different student cohorts identified by analysis of student learning data. See [80] for a more detailed explanation of the figure.

Regarding the internal characteristics of the tool (factor I), each online learning module consists of an assessment component with just one or two questions and an "instruction and practice" component that consists of text and practice problems focused on enabling students to solve the problems in the assessment component. Students are given a total of five attempts on the assessment of each module but are required to make a first attempt on the assessment problems before being allowed to study the learning resources, a design that is inspired by research on "preparation for future learning" and productive struggle, which emphasize the importance of struggling with problems before being provided support [81]. Each module focuses on one topic, and a sequence of 7–12 modules cover the content of one or two weeks of a typical course. One can roughly think of online learning modules as blending the contents of a textbook with problems from the end-of-chapter quiz according to sub-topics (See Figure 2A for a visual illustration).

The assessment components of each digital module serve as a formative self-assessment tool. The assessment allows students to monitor their learning and be able to self-regulate better based on their level of background preparation (factor II). Students with higher incoming knowledge may quickly pass the easier modules at the beginning of the sequence and start interacting with more difficult content toward the end; on the other hand, those who have less incoming knowledge will have ample opportunity to engage with more basic learning modules at the beginning of the sequence. This unique experience is not possible in the traditional textbook (or digital textbook) format, where instructional text and assessment problems are separated, and assessment follows instruction.

In one study analyzing students' interactions with online learning modules in a university-level introductory physics course, the data show that students spontaneously use different strategies to engage with online learning modules, based on their incoming knowledge and assessment outcomes [80]. As shown in Figure 2B, the analysis of data for how students engaged with these self-paced modules identified four clusters of students, with each cluster adopting a different strategy working through a sequence of eight learning modules. The strategy used by students in each cluster is represented in one of the four bar graphs shown in Figure 2B. Each vertical bar in a bar graph corresponds to one learning module in the sequence, and different colors represent different types of strategies that students adopted to engage with each module. For example, purple represents students who pass a given module on their first one or two attempts without needing to access the instructional content. As shown in the figure, almost every student in cluster 1–1 passed modules M1 and M2 on their initial attempt, and the majority of students passed all of the modules without interacting with the instructional materials. The students in that cluster likely have high incoming knowledge of relevant content. On the other hand, students who passed a given module after accessing the instructional content are represented using orange. As can be seen, most students in cluster 1–4 needed to study the instructional materials starting from module M-1, whereas students in cluster 1–2 could pass the first couple of modules but are much more likely to study the instructional materials toward the end of the sequence. The red color represents students who either did not pass the module after studying the content or made multiple very short guessing attempts before passing. This mode of passing is seen on the last module of the sequence among students in clusters 1–2 and 1–4. It is worth noting that students in cluster 1–2 seem to have changed their engagement strategy between the first and last module in the sequence, which can be seen as a sign of active self-regulation.

Another example that shows the importance of considering students' internal characteristics, e.g., their prior knowledge and skills, in the design of digital learning tools, came from an online tutorial first implemented in the form of a sequence of online learning modules [82], as illustrated in Figure 3.

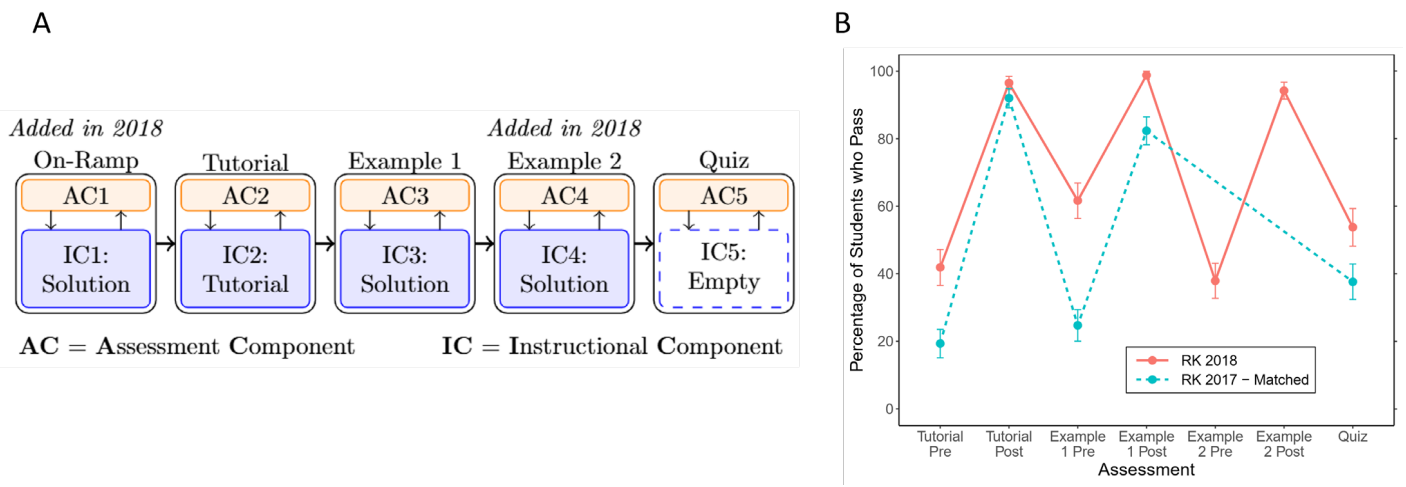

**Figure 3.** Tutorial in the form of online learning modules. (**A**) The online tutorial design. Modules with dashed lines are added in the second semester. (**B**) Comparison of student performance between the two semesters.

The initial tutorial, designed based on the work of [70], consists of three online learning modules, a Tutorial module, an Example module, and a Quiz module, which students are required to complete in the given sequence as part of their homework. The instructional component of the "Tutorial" module contains a step-by-step tutorial walking students through the process of solving a challenging introductory-level physics problem. The assessment component of the "Example" module contains a similar problem, and the instructional component contains a worked-out solution of the problem. In both the tutorial and the example module, students were required to make one initial attempt at the problem before accessing either the tutorial or the worked-out solution. In the last module (the Quiz Module), students were presented with another similar problem but with no instructions available.

The fraction of students who could solve the assessment problem on each module within the first three attempts is plotted in cyan dots connected by the dashed line in Figure 3B. The dots labeled "Tutorial Pre", and "Example 1 Pre" represent the fraction of correct answers prior to accessing the instructional materials, and the "Tutorial Post" and "Example 1 Post" represent the fraction of correct answers after students have accessed the learning materials. Notably, while on the "Tutorial" module, the "Post" attempt correct rate is close to 80%, on the "Example Pre" attempt, the correct rate dropped back to between 20 and 30%. A similar pattern is observed between "Example Post" and "Quiz". Those observations indicate that while most students can learn to solve the exact same problem after accessing either the tutorial or the worked-out solution, as demonstrated by the high percentage on "Post", they have limited ability to transfer their learning to a somewhat different problem context even though the underlying physics principles are the same.

Based on those results, in the next semester, two new modules were added to the tutorial sequence, as shown in Figure 3A. The first is an "on-ramp" module that contains two problems that are simpler than the problem in the first tutorial module. The addition of the "on-ramp" module is based on the consideration that many students may need an opportunity to either learn or be reminded of basic problem-solving steps before they are ready to learn from the more complicated problem in the tutorial module (considerations of both factors I and II in the SELF framework). The second module is the "example 2" module, which is designed in a similar fashion as the "example 1" module but with a different problem. This module provides students with another practice opportunity before they access the last module ("Quiz") in the sequence.

Students' performance on the updated tutorial sequence is plotted in Figure 3B in red dots and a solid line. Their performance on the on-ramp modules is not shown since it is not part of the comparison. As seen in the figure, significant performance improvements are observed for "Tutorial Pre", "Example 1 Pre", and "Example 1 Post". Notably, the largest improvement was observed for the "Example 1 Pre" attempt, which took place after students finished the Tutorial module and before accessing the worked solution in the "Example 1" module. This indicates that with more preparation of basic skills from the on-ramp module, students are more likely to transfer what they have learned to similar new problems.

*4.2. Considering the External Characteristics of Digital Learning Tools to Support Student Self-Regulation (Factors III and IV)*

While carefully developed adaptive interactive learning tools that account for students' prior knowledge and different backgrounds are more likely to engage students in the learning process, our research shows that if students are not adequately supported and incentivized, they may not take advantage of the self-paced digital tools to learn in an effective manner, thus, the tools will not necessarily help them learn. In fact, the top two quadrants of the SELF framework in Figure 1 (internal characteristics of the learning tools and students) are often considered in the development of evidence-based digital learning tools; however, the equally important lower two quadrants of SELF focusing on effective implementation of the tools and ensuring that students receive encouragement and support from instructors and peers to self-regulate and engage meaningfully with the tool are often given less attention. These lower two quadrants are likely to play a critical role in whether students, especially those with lower prior knowledge and self-efficacy, take advantage of them.

The lower left quadrant of the SELF framework (factor III) pertains to external learning tool characteristics for self-regulated learning, i.e., how the tools are implemented and incentivized to optimize student motivation to engage with them as part of a course. Consideration of the various types of support in factor III during the implementation of the learning tools is critical to ensure that most students are motivated to engage with them effectively. For example, to help students engage effectively with the self-paced digital learning tools, an instructor could incentivize participation in learning via better grade incentives to ensure that students work on them as prescribed. A good example showing how a lack of proper grade incentives can negatively impact student learning even with well-designed digital learning tools comes from our interviews with students about their experiences learning from the course videos in flipped courses [83].

The flipped or hybrid course model has been widely implemented at many colleges and universities, especially during the last decade. This approach requires an instructor to completely move away from the traditional "sage on the stage" role and instead devote class time to interacting with students more as a coach to help with problem-solving or other concrete skill development while developing a robust knowledge structure. In a typical flipped classroom setting, students first engage with content through instructional materials such as video lectures, textbooks, or reading assignments. In our interviews with 37 students taking college introductory physics courses at the end of the fall 2020 semester when all courses were being taught remotely due to the COVID-19 pandemic [83], we found overwhelmingly that when students had no grade incentive for doing out-of-class or in-class activities in a flipped course, and if the solutions to all problems and in-class assignments were posted on the course website after each class, many students had difficulty in self-regulating and simply stopped doing the weekly out-of-class and in-class activities early on in the semester. They only watched the relevant videos and browsed through all assessments right before each exam. They admitted that the lack of a grade incentive led them to deprioritize the weekly course activities. In other words, a negative feedback loop occurred: students quickly realized that going to the class was "useless" if they had not completed the out-of-class work involving watching videos and engaging with the corresponding assessments because the instructor immediately put them in a Zoom breakout room and expected them to discuss their answers with their peers in small groups. A handful of students who had completed the out-of-class work dominated the discussion, while the unprepared majority were disengaged. Since prior engagement with video lectures was expected, the general discussions after each peer discussion in the Zoom breakout rooms were not productive for them, either.

In essence, the implementation of digital tools such as course videos in flipped courses without proper incentivization mechanisms such as grade incentives to motivate meaningful engagement devolved the course into a learning experience akin to taking an asynchronous massive open online course (MOOC) [84] but with a few high-stakes synchronous exams for most students [83]. Many students did not keep up with the course videos and other course materials and crammed a few days before each exam, something that has been shown to be detrimental to long-term retention in some studies [85]. The lack of engagement and learning did not stem from a lack of effort on the part of the instructors, who in good faith spent a tremendous amount of time and effort to make customized flipped videos for their course [83]. Rather, consistent with the SELF framework, it resulted from the implementation of the different components of the course without grade incentives (consistent with factor III in the framework), which was not conducive to effective engagement. For most students, keeping a consistent and high level of engagement with the course content over an extended period can be challenging due to the high level of self-regulation that is required. By contrast, most students found flipped courses in a remote format that were otherwise similar yet had a grade incentive to be effective, and they kept up with those courses [83]. This sentiment of the students is consistent with our prior research, which shows that a majority of students have difficulty engaging with online evidence-based self-paced learning tools unless there is a grade incentive associated with out-of-class engagement with online tools [68,70]. It should be pointed out that properly designed grade incentives are just one of the many aspects in the implementation of digital learning tools, such as online videos in a flipped course, that can serve as an effective scaffold that continues to motivate students. As discussed later in this section, other aspects such as student self-efficacy [3] must also be accounted for appropriately in the effective implementation of digital learning tools consistent with the SELF framework.

Interviewed students also noted that some instructors had made their courses completely asynchronous with video lectures and had turned their allocated class time into office hours [83]. All that counted toward a grade in their course were a few exams. Many students confessed that they did not go to the office hours because they did not keep up with the video lectures and crammed the material right before the exams [83]. The interviewed students were in general very dissatisfied with their learning in the asynchronous courses due to the manner in which the course modules were implemented (consistent with factor III of the SELF framework).

In a different study [68,70], students were given opportunities to work through web-based adaptive tutorials in introductory physics courses outside of class as self-study tools. Students were provided these optional tutorials after traditional instruction in relevant topics and were then given quizzes that included problems identical to the tutorial problems regarding the physics principles involved but had different contexts. We find that many students who worked through the tutorials as self-study tools engaged with them superficially and struggled to transfer their learning to solve quiz problems that used the same physics principles. On the other hand, students who worked on the same web-based tutorials in supervised, one-on-one situations performed significantly better than them regardless of their physics grade in the course at the time they engaged with the web-based tutorial in a supervised manner. These empirical research findings suggest that many introductory physics students may not engage effectively with self-paced digital learning tools unless they are provided additional incentives and support to aid with self-regulation consistent with factor III in the SELF framework in Figure 1.

On the other hand, a positive example of using grade incentives to encourage productive self-regulation with digital learning tools among students (consistent with factor III of the framework in Figure 1) is using extra credit to encourage work-planning and reduce due-date cramming, shown in Figure 4.

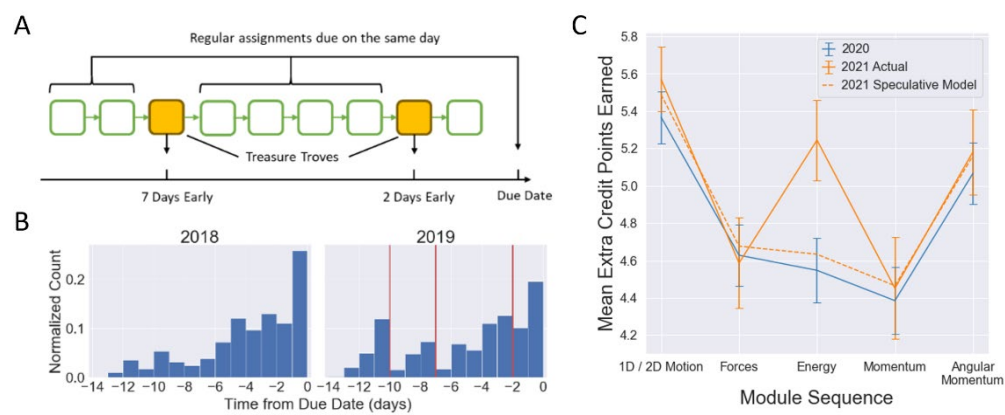

**Figure 4.** Improving students' work distribution using credit and non-credit incentives. (**A**) Schematic illustration of treasure troves. (**B**) Comparison of the start of study sessions before and after the implementation of treasure troves. The assignment is due on day 0. The red lines indicate dates on which extra credit incentives were provided. (**C**) Change in average extra credit points earned before and after implementation of "nudging" survey. Blue line shows the data before implementation, orange dashed line shows the best estimate based on previous year's data, and orange solid line shows the actual measured extra credits earned after the implementation of "nudging" on the Energy module's sequence.

As reported by Felker et al. [86], when assigned to complete a sequence of 10 online homework assignments over a period of 14 days, many students chose to start all of the assignments close to the due date and spend a relatively short time working on each assignment. To encourage better work distribution, the researchers implemented extra credit incentives in the form of "treasure trove" quizzes. Each treasure trove quiz contains only one question: "Would you like to claim your extra credit for completing the assignments early?", and students earn some extra course credit if they answer yes. As illustrated in Figure 4A, treasure trove assignments are embedded into a sequence of multiple online homework assignments that are due on the same day, and students can only access each treasure trove after they finish all the preceding assignments. However, the due date of each treasure trove is set to several days before the assignment due date, so only when students finish all preceding assignments earlier than the due date can they obtain an extra credit assignment.

To illustrate how incentivizing students to engage with digital learning tools consistent with factor III in the framework increases student engagement, in Figure 4B, we plot the fraction of study sessions started each day for a homework assignment sequence and compare the data before and after adding the treasure troves. A study session is defined as a period of continued engagement with the online homework by a student. As shown in Figure 4B, in 2018, which was before the implementation of treasure troves, most study sessions tended to be started either 2–4 days before or on the same day as the assignment due date (day 0 on the figure). After adding the treasure troves in 2019, there was a clear shift of study sessions toward earlier dates and concentrated on days −10, −7, and −2, which were the treasure trove due dates. Further analysis in the paper also found that both high-performing and low-performing students are equally likely to take advantage of the treasure trove opportunities. In their course evaluation, some students noted that the extra credit simply gave them a reason to start the assignment early.

While credit incentives can be easy and effective to help students engage with self-paced digital learning tools, they are far from the only external implementation factor (factor III of the framework in Figure 1) that could facilitate students' self-regulation. In some cases, non-credit incentives such as "nudging" could be highly effective methods for motivating students. For example, in the same introductory physics course administered in 2020 and 2021, students were asked to complete a survey on their intention and plan to obtain extra credits for one homework sequence. Students were in no way obliged to follow their stated plan, nor was there any requirement for the quality or level of detail of their plan. However, when comparing the average amount of extra credit points acquired for each homework sequence with students from the previous semester, we were surprised to find that students who answered the survey acquired significantly more extra credit points on that particular homework sequence. In Figure 4C, we show the amount of extra credit earned each week. The dashed orange line represents the best prediction of extra credit that students would have earned each week, based on data from the previous year (blue line), and the estimated year-to-year difference (see Felker 2023 for details of the modeling process [86]). The solid orange line shows the actual amount of extra credit earned by students, who were nudged to write a plan during the week labeled "Energy". In other words, simply asking students to make a plan for completing the assignments (with no grade incentive) was enough to encourage more students to actually complete assignments earlier than expected [86]. This serves as an example of how factor III of the framework could be integrated into the implementation of digital tools without using grades as an incentive.

*4.3. Adding the Human Touch to Digital Learning Tools (Factor IV)*

The lower right quadrant of the SELF framework, or factor IV, focuses on characteristics related to student–environment interaction, e.g., how students interact with their surroundings, how they manage their time, and how they regulate themselves. Factor IV also includes community support, e.g., support students may receive from their environments such as help from the instructors, peers, family, and advisors to manage their time better and engage in learning effectively. In particular, students' engagement with digital learning tools can be impacted by how students self-regulate and manage their time, which can further be impacted by human elements including whether students feel like they are part of the classroom community. Furthermore, encouragement and support from others to engage with the tools can enhance their self-efficacy. Therefore, a critical consideration in scaling online education is humanizing it and ensuring that students feel like they are part of the classroom community. Students want to feel supported by human beings, and they thrive when they feel a bond with their instructors and peers. Building the classroom community and facilitating effective interactions among students and instructors do not necessarily require instructors to spend a lot of time. Our research involving interviews with 37 students during the online classes of the pandemic suggests that what matters is the genuine positive intent and effort [83]. Instructors will be surprised at how much students are touched by their small gestures and time commitments to empathetic discussions as well as personal support in online courses. In other words, when instructors spend a small amount of time connecting meaningfully with students and creating a classroom community, it can incentivize students to prioritize course content even in remote courses, reduce procrastination and time-management issues, and increase their attention and engagement (consistent with factor IV in the framework in Figure 1).

Our interviews with 37 students in the fall of 2020 semester during the COVID-19 period [83] suggest that while access to digital technology made online instruction possible during the pandemic, the most important factors that were tied to student satisfaction and performance were those related to human connections with instructors and peers. Students reported that in courses with a synchronous component, they appreciated the opportunities to be part of the classroom community, especially during the synchronous components of the course but also via asynchronous Q and A platforms such as Piazza. This humanistic feeling of being part of the classroom community had several advantages. The social aspects of such courses, including the opportunity for interactions and communications with the instructors and peers, reduced procrastination, improved time-management and self-regulation, and increased student attention and engagement in the course when they participated. Students were so starved for human interaction that they overwhelmingly pointed out even minute efforts made by instructors to inject some human element into their interactions, such as spending a few minutes at the beginning of their classes checking in with them for student well-being and asking how they were doing and even talking to them about their own adversity during the challenging times. Thus, these empathetic discussions over Zoom were usually quite short but they had an outsized impact on student morale. Some students noted that they wished that this type of humanistic connection with instructors and peers at the beginning of each class was the norm, even in regular courses during normal times. Students did not want to miss the classes in which instructors incorporated these short check-ins with students as a regular feature.

Asynchronous communication platforms such as Slack and Microsoft Teams, which provide instant messaging and threaded conversation, can be a highly useful tool for "humanizing" communication with students and building the sense of community in a physics course. Compared to the traditional email, these messaging platforms significantly lower the barrier of communication by reducing the (often implicit) need for formalism and instilling a sense of informal and casual dialogue. Put in simpler terms, the use of these platforms speeds up communication by saving a large amount of "Dear professor/Student" and "Sincerely" in the dialogue. Furthermore, as shown in Figure 5, the use of emojis and animated GIFs during threaded messaging facilitates the expression of emotion and adds a human touch to the conversation, where a "perfect, thank you" email can now be simply replaced by a thumbs up or a heart emoji. For the majority of the current generation of students who grew up immersed in social media and instant messaging, this type of communication feels very natural and comfortable.

In one instance that took place during an introductory physics class in the fall of 2020 during the COVID-19 period, many students experienced a widespread network outage on the day of an online mid-term exam due to a construction accident near a student dorm. The instructor of the course was immediately aware of the scale and nature of the disruption by glancing through the large number of instant messages received on MS Teams, and quickly made an announcement in Teams to extend the due date of the exam by one day (Figure 5). As a result, several students expressed their appreciation for the quick response via emojis and figures in the same threaded post.

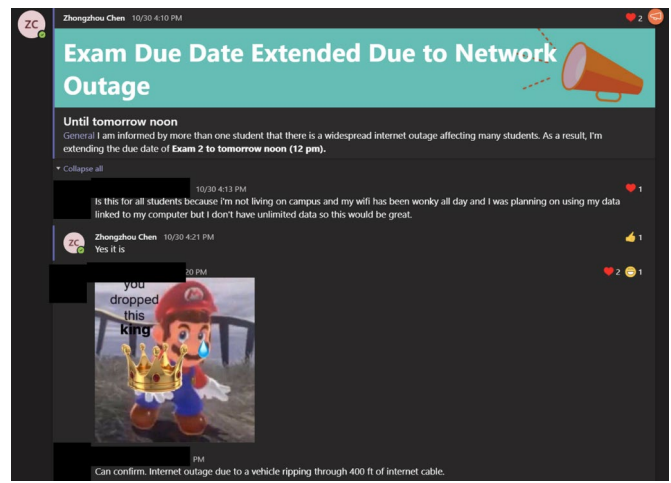

**Figure 5.** Screenshot of online communication on Microsoft Teams (Version 1.3.00.4461) from the fall 2020 semester during the COVID-19 period showing an announcement by the instructor extending the exam due date, and reactions from students in the form of emojis and internet memes.

Whether performed as part of in-person, remote, or hybrid courses, humanizing digital learning will increase student enthusiasm and commitment to be part of the community, their sense of belonging, and concentration to participate regularly. Interactions with instructors and peers can help students engage meaningfully with the course materials and manage their time better. Instructors should have a synchronous component in their courses, as in the flipped courses, and incentivize students with many low-stakes grade incentives interspersed throughout the course. However, if the entire course must be asynchronous, then asynchronous platforms can play a central role in humanizing learning and increasing student interactions with instructors and peers. This can ensure that students feel like a part of the community and self-regulate and engage in learning from digital tools productively, consistent with the SELF framework for guiding the design/curation and implementation of digital learning.

*4.4. Usefulness of SELF Framework for Harnessing Generative AI for Learning*

For the past few years, the excitement around generative AI, in particular LLM-based services such as ChatGPT, has led to increased interest in effectively using digital learning pedagogically to improve learning at all levels, e.g., see [34–47]. While the emergence of generative AI presents both unprecedented opportunities and challenges for education, the SELF framework proves particularly valuable for guiding the implementation of GenAI tools because it provides a structured approach to addressing the unique challenges this technology presents while maximizing its potential benefits.

For example, our prior research suggests [83] that the effectiveness of digital learning tools will not go as far as many people have envisioned in the past unless the human element is woven into how to integrate and implement these tools in a meaningful way, keeping in mind all four dimensions of the SELF framework. In particular, generative AI has great potential to enhance learning, including personalizing it and making it suited for students' needs if one keeps in mind, e.g., the internal and external characteristics of the students and the learning environments to ensure that a variety of students from diverse backgrounds can benefit from it [87,88]. Ensuring self-regulation is essential for maximizing the potential of generative AI to personalize learning for diverse student groups. Thus, to achieve this, all four aspects of the SELF framework must be considered holistically, including the design and implementation of learning tools while accounting for students' internal and external characteristics.

The internal characteristics of GenAI tools demand careful consideration in educational settings. While these tools offer powerful capabilities for personalized feedback and adaptive learning, they also present risks of hallucinations and misinformation [89,90]. Recent studies have demonstrated varying accuracy across different types of questions [34,38], and while some research shows promise for enhanced learning experiences through AI-generated feedback [40,41], other studies reveal the potential for compromised learning outcomes [36]. Understanding these affordances and limitations can help educators design appropriate learning activities that leverage AI's strengths while mitigating its weaknesses. In particular, while research on the use of generative AI to improve student learning in STEM disciplines has been relatively limited until now, its use is likely to grow rapidly. Researchers have also looked into using generative AI to grade [40,91,92] students' written responses, and generate personalized feedback based on each student's written response to questions, which combines factors I (tool characteristics) and II (student characteristics) of the SELF framework. Some early studies have shown that students sometimes prefer AI-generated feedback over human-generated [41]. Education researchers and educators are also considering ways in which students can engage with ChatGPT effectively by asking good questions and learn the basics of prompt engineering [93] to enhance and personalize student learning.

There is also growing interest in considering restricting the types of information from which the generative AI model can draw from when answering questions, to minimize the negative impact of misinformation and bias in online materials not properly vetted for student learning. These could be achieved via techniques such as fine-tuning [94,95] and Retrieval Augmented Generation (RAG) [96]. For example, Khan Academy is using generative AI in a way that allows the language models to base their answers on vetted databases and materials within Khan Academy [97]. If a student attempts to work on a tutorial on that platform and makes certain types of mistakes in the assessment tasks, then based upon their mistakes, the generative AI may recommend that they work on other more basic tutorials relevant to their topic of interest to strengthen their basics before coming back to their original tutorial. Similarly, there is growing interest in understanding the impact of generative AI hallucinations on student learning. For example, temperature is one of several parameters that can impact the level of creativeness of the model [98]. In particular, while generative AI can potentially be leveraged to foster creativity among students, it can also produce answers that are misleading. If the students have not been given the opportunity to develop the ability to recognize responses that are misinformed or biased, they are likely to be misled. Thus, investigating pedagogical approaches to helping students develop the ability to distinguish between correct and incorrect feedback (as well as biased feedback) from generative AI would be valuable for enhancing critical thinking skills. Researchers and educators are also contemplating how to help students develop skills so that they can ask generative AI good questions and how to make the dialogue between the generative AI and students more Socratic, e.g., for each question that a student asks the generative AI, the generative AI first answers the question, and then asks the student a related question to reflect upon. This type of interaction between the generative AI and students is reminiscent of reciprocal teaching approaches, such as those promoted by Palincsar and Brown [99] and adapted in physics, e.g., via the Personal Assistant to Learning (PAL) tutor [100,101]. Investigations of student learning in controlled studies involving different interventions such as these can be valuable for harnessing the true potential of generative AI in enhancing and personalizing student learning.

On the other hand, many researchers and instructors are concerned about the potential negative impact that generative AI could have on the development of students' problem-solving, reasoning, and meta-cognitive skills. For example, one concern is that students can receive too much help from generative AI on their take-home assessments or practice tasks, such as using generative AI for writing their essays and computer codes and solving their mathematics and physics problems [36]. Since class time is limited, self-paced homework plays a key role in helping students learn both the content and skills and develop self-reliance in being able to solve problems. However, generative AI services such as ChatGPT could significantly compromise the learning experience and thwart learning and knowledge organization as well as skill development by generating complete solutions for the student [102]. Similarly, even when students only use GenAI for reference rather than obtaining solutions during practice, there is still the risk of them being exposed to hallucinations, misinformation, and bias inherent to LLMs [103–105]. We believe that this is a case where the tool (generative AI) is being implemented in learning, utilizing only its internal character (factor I) but neglecting all other three factors in the SELF framework. In other words, generative AI has not been integrated into the learning experience with proper consideration of students' internal and external characteristics in its implementation.

We would also like to argue that by fully considering all four factors of the SELF framework, including the affordances and constraints of generative AI, the above-mentioned situations could actually present new and valuable opportunities for creatively developing key skills for students. For example, instructors could task generative AI to generate the initial computer coding for a specific task, embedding specific types of coding errors within it [106], then use it as an opportunity for students to learn to debug either individually or as a group. A second GenAI assistant could then be tasked to either grade students' reasoning on the debug task, or provide personalized feedback and hints during the process. In this example, GenAI as a tool is an integral part of the learning experience aiming at learning to debug (factor III), while providing students with the opportunity and assistance to practice reflection individually (factor II) or as a group (factor IV). Similarly, while asking generative AI to write essays for students would be unethical and thwart learning [102], instructors could design a task in which generative AI writes the same essay from multiple perspectives and ask the students to either critically assess those perspectives, synthesize the different perspectives, or ask students to combine their own unique background and experience with the draft writing provided by generative AI.

Investigations related to other issues such as how to get students to collaborate meaningfully and learn while having generative AI participate can also be valuable [107]. Other investigations that focus on how to help students develop critical thinking and creative thinking skills [108,109] can be useful in this regard. Furthermore, these types of issues can be valuable not only for developing critical thinking skills but also for training in ethics.

In summary, by utilizing the SELF framework to ensure that students with different internal and external characteristics can be adequately supported and by tailoring the internal and external characteristics of the learning environments, researchers and educators will be able to harness the power of generative AI and digital technology productively, safely, and creatively to not only overcome the potential concerns but also greatly enhance students' learning experience and learning outcomes.

## 5. Summary

The educational crisis caused by the COVID-19 pandemic in 2020 demonstrated that considering evidence-based approaches to harness digital learning is essential. Digital learning tools, from basic online homework systems to sophisticated AI assistants, have become integral to higher education. As we look to the future, digital learning tools, including emerging technologies like generative AI, will likely become increasingly integrated into both online and in-person courses.

The SELF framework provides a comprehensive approach for maximizing the benefits of these tools while minimizing potential pitfalls. Our research demonstrates that effective implementation requires attention to tool design, implementation strategies, student characteristics, and community support. In particular, the effectiveness of digital learning tools depends on careful consideration of all four dimensions of the SELF framework. First, the tools themselves must be designed with evidence-based adaptive features that accommodate diverse learners. Second, proper implementation with appropriate grade incentives and support structures is essential—as shown by our studies of online tutorials and flipped classrooms. Third, careful attention must be paid to students' internal characteristics, including their prior knowledge and self-regulation abilities. Finally, and equally critically, student external characteristics must be considered, and the human elements must be thoughtfully woven into digital learning implementations. This includes building the classroom community, facilitating meaningful peer interactions, and providing instructor support based upon student internal and external characteristics—factors that our research shows are vital for helping students engage productively with digital tools. By holistically considering these four dimensions, educators can better harness digital learning tools to create more equitable and inclusive learning environments that support students from diverse backgrounds and prior preparations. This comprehensive approach will be especially important as institutions continue integrating new technologies into their instructional practices while maintaining a focus on evidence-based pedagogical principles that promote effective learning for all students.

Furthermore, given its potential, the use of generative AI in personalizing and enhancing learning is likely to show tremendous growth in the coming decades. In the era of generative AI, researchers and educators should investigate effective approaches to assessing student learning, including how to ensure that student learning is maximized through self-paced digital homework assignments and tasks that productively utilize generative AI. In the future, even in-person courses are increasingly likely to use self-paced digital learning tools, including those involving generative AI. Research involving generative AI along different dimensions of the SELF framework, including those suggested in Section 4, can go a long way to harnessing the potential of digital learning and generative AI to personalize learning for students with diverse backgrounds.

Evidence-based adaptive self-paced digital learning tools when incentivized appropriately will go a long way in helping students from diverse backgrounds and prior preparations excel and can be central in supporting equitable and inclusive learning in an era of generative AI. Without leveraging these self-paced digital learning tools, which are likely to be increasingly enhanced by generative AI, it is difficult for educators to level the playing field due to the difficulty in designing in-class instruction alone to cater to students with diverse prior preparations, keeping in mind that students with systemic advantages are more likely to have higher levels of prior preparation. Providing students with appropriate incentives and support is vital for ensuring that instructors' efforts in bolstering student learning by incorporating self-paced digital learning tools in their instructional design bear fruit so that students from diverse prior preparations learn the content and develop critical thinking skills by engaging with digital learning tools as part of their course.

**Author Contributions:** Conceptualization, C.S. and Z.C.; resources, Z.C. and C.S.; writing—original draft preparation, C.S. and Z.C.; writing—review and editing, C.S. and Z.C.; visualization, Z.C. All authors have read and agreed to the published version of the manuscript.

**Funding:** This research was funded by National Science Foundation grant number DUE-1845436, and University of Central Florida Digital Learning Course Reform and Innovation Fund.

**Conflicts of Interest:** The authors declare no conflict of interest.

## References

1. Dewey, J. *Democracy and Education*; Macmillan: New York, NY, USA, 1916.
2. Vygotsky, L. *Mind in Society: The Development of Higher Psychological Processes*; Harvard University Press: Cambridge, MA, USA, 1978.
3. Bandura, A. Self-efficacy: Toward a unifying theory of behavioral change. *Psychol. Rev.* **1977**, *84*, 191–215. https://doi.org/10.1037/0033-295X.84.2.191.
4. Green, C.; Brewe, E.; Mellen, J.; Traxler, A.; Scanlin, S. Sentiment and thematic analysis of faculty responses: Transition to online learning. *Phys. Rev. Phys. Educ. Res.* **2024**, *20*, 010151. https://doi.org/10.1103/PhysRevPhysEducRes.20.010151.
5. Li, Q.; Xu, S.; Xiong, Y.; He, W.; Zhou, S. Comparing the perception of emergency remote teaching experience between physics and nonphysics students. *Phys. Rev. Phys. Educ. Res.* **2023**, *19*, 020131. https://doi.org/10.1103/PhysRevPhysEducRes.19.020131.
6. Beardsley, M.; Albó, L.; Aragón, P.; Hernández-Leo, D. Emergency education effects on teacher abilities and motivation to use digital technologies. *Br. J. Educ. Technol.* **2021**, *52*, 1455–1477.
7. Qureshi, M.I.; Khan, N.; Raza, H.; Imran, A.; Ismail, F. Digital technologies in education 4.0. Does it enhance the effectiveness of learning? *View Item* **2021**, *15*, 31.
8. Jevsikova, T.; Stupurienė, G.; Stumbrienė, D.; Juškevičienė, A.; Dagienė, V. Acceptance of distance learning technologies by teachers: Determining factors and emergency state influence. *Informatica* **2021**, *32*, 517–542.
9. Shohel, M.M.C.; Ashrafuzzaman, M.; Islam, M.T.; Shams, S.; Mahmud, A. Blended teaching and learning in higher education: Challenges and opportunities. In *Handbook of Research on Developing a Post-Pandemic Paradigm for Virtual Technologies in Higher Education*; IGI Global: Hershey, PA, USA, 2021; pp. 27–50.
10. Williamson, B.; Eynon, R.; Potter, J. Pandemic Politics, Pedagogies and Practices: Digital Technologies and Distance Education During the Coronavirus Emergency. *Learn. Media Technol.* **2020**, *45*, 107–114.
11. Giovannella, C. Effect induced by the COVID-19 pandemic on students' perception about technologies and distance learning. In *Ludic, Co-Design and Tools Supporting Smart Learning Ecosystems and Smart Education: Proceedings of the 5th International Conference on Smart Learning Ecosystems and Regional Development*; Springer: Singapore, 2021; pp. 105–116.
12. Bergdahl, N.; Nouri, J. COVID-19 and crisis-prompted distance education in Sweden. *Technol. Knowl. Learn.* **2021**, *26*, 443–459.
13. Mhlanga, D.; Denhere, V.; Moloi, T. COVID-19 and the Key Digital Transformation Lessons for Higher Education Institutions in South Africa. *Educ. Sci.* **2022**, *12*, 464.
14. Meletiou-Mavrotheris, M.; Eteokleous, N.; Stylianou-Georgiou, A. Emergency Remote Learning in Higher Education in Cyprus during COVID-19 Lockdown: A Zoom-Out View of Challenges and Opportunities for Quality Online Learning. *Educ. Sci.* **2022**, *12*, 477.
15. Barlovits, S.; Jablonski, S.; Lázaro, C.; Ludwig, M.; Recio, T. Teaching from a Distance—Math Lessons during COVID-19 in Germany and Spain. *Educ. Sci.* **2021**, *11*, 406.
16. Sofianidis, A.; Meletiou-Mavrotheris, M.; Konstantinou, P.; Stylianidou, N.; Katzis, K. Let students talk about emergency remote teaching experience: Secondary students' perceptions on their experience during the COVID-19 pandemic. *Educ. Sci.* **2021**, *11*, 268.
17. Camilleri, M.A.; Camilleri, A.C. The acceptance of learning management systems and video conferencing technologies: Lessons learned from COVID-19. *Technol. Knowl. Learn.* **2022**, *27*, 1311–1333.
18. McKagan, S.B.; Perkins, K.K.; Dubson, M.; Malley, C.; Reid, S.; LeMaster, R.; Wieman, C.E. Developing and researching PhET simulations for teaching quantum mechanics. *Am. J. Phys.* **2008**, *76*, 406–417.
19. Wieman, C.E.; Adams, W.K.; Perkins, K.K. PhET: Simulations that enhance learning. *Science* **2008**, *322*, 682–683.
20. Belloni, M.; Christian, W.; Brown, D. Open source physics curricular material for quantum mechanics. *Comput. Sci. Eng.* **2007**, *9*, 24—31.

21. Kohnle, A.; Douglass, M.; Edwards, T.J.; Gillies, A.D.; Hooley, C.A.; Sinclair, B.D. Developing and evaluating animations for teaching quantum mechanics concepts. *Eur. J. Phys.* **2010**, *31*, 1441–1455.
22. Park, S.I.; Lee, G.; Kim, M. Do students benefit equally from interactive computer simulations regardless of prior knowledge levels? *Comput. Educ.* **2009**, *52*, 649–655. https://doi.org/10.1016/j.compedu.2008.11.014.
23. Canright, J.P.; White Brahmia, S. Modeling novel physics in virtual reality labs: An affective analysis of student learning. *Phys. Rev. Phys. Educ. Res.* **2024**, *20*, 010146. https://doi.org/10.1103/PhysRevPhysEducRes.20.010146.
24. Osadchyi, V.; Valko, N.; Kuzmich, L. Using augmented reality technologies for STEM education organization. *J. Phys. Conf. Ser.* **2021**, *1840*, 012027.
25. Marks, B.; Thomas, J. Adoption of virtual reality technology in higher education: An evaluation of five teaching semesters in a purpose-designed laboratory. *Educ. Inf. Technol.* **2022**, *27*, 1287–1305.
26. Mystakidis, S.; Christopoulos, A.; Pellas, N. A systematic mapping review of augmented reality applications to support STEM learning in higher education. *Educ. Inf. Technol.* **2022**, *27*, 1883–1927.
27. Singh, R.P.; Javaid, M.; Kataria, R.; Tyagi, M.; Haleem, A.; Suman, R. Significant applications of virtual reality for COVID-19 pandemic. *Diabetes Metab. Syndr. Clin. Res. Rev.* **2020**, *14*, 661–664.
28. Mehrfard, A.; Fotouhi, J.; Taylor, G.; Forster, T.; Armand, M.; Navab, N.; Fuerst, B. Virtual reality technologies for clinical education: Evaluation metrics and comparative analysis. *Comput. Methods Biomech. Biomed. Eng. Imaging Vis.* **2021**, *9*, 233–242.
29. Haleem, A.; Javaid, M.; Qadri, M.A.; Suman, R. Understanding the role of digital technologies in education: A review. *Sustain. Oper. Comput.* **2022**, *3*, 275–285. https://doi.org/10.1016/j.susoc.2022.05.004.
30. Dreimane, S.; Upenieks, R. Intersection of Serious Games and Learning Motivation for Medical Education: A Literature Review. In *Research Anthology on Developments in Gamification and Game-Based Learning*; I.R. Management Association, Ed.; IGI Global: Hershey, PA, USA, 2022; pp. 1938–1947.
31. Nkomo, L.M.; Daniel, B.K.; Butson, R.J. Synthesis of student engagement with digital technologies: A systematic review of the literature. *Int. J. Educ. Technol. High. Educ.* **2021**, *18*, 34.
32. Abriata, L.A. How technologies assisted science learning at home during the COVID-19 pandemic. *DNA Cell Biol.* **2022**, *41*, 19–24.
33. Polly, D.; Martin, F.; Guilbaud, T.C. Examining barriers and desired supports to increase faculty members' use of digital technologies: Perspectives of faculty, staff and administrators. *J. Comput. High. Educ.* **2021**, *33*, 135–156.
34. Kortemeyer, G. Could an artificial-intelligence agent pass an introductory physics course? *Phys. Rev. Phys. Educ. Res.* **2023**, *19*, 010132. https://doi.org/10.1103/PhysRevPhysEducRes.19.010132.
35. Dahlkemper, M.N.; Lahme, S.Z.; Klein, P. How do physics students evaluate artificial intelligence responses on comprehension questions? A study on the perceived scientific accuracy and linguistic quality of ChatGPT. *Phys. Rev. Phys. Educ. Res.* **2023**, *19*, 010142. https://doi.org/10.1103/PhysRevPhysEducRes.19.010142.
36. Kortemeyer, G.; Bauer, W. Cheat sites and artificial intelligence usage in online introductory physics courses: What is the extent and what effect does it have on assessments? *Phys. Rev. Phys. Educ. Res.* **2024**, *20*, 010145. https://doi.org/10.1103/PhysRevPhysEducRes.20.010145.
37. Küchemann, S.; Steinert, S.; Revenga, N.; Schweinberger, M.; Dinc, Y.; Avila, K.E.; Kuhn, J. Can ChatGPT support prospective teachers in physics task development? *Phys. Rev. Phys. Educ. Res.* **2023**, *19*, 020128. https://doi.org/10.1103/PhysRevPhysEducRes.19.020128.
38. Polverini, G.; Gregorcic, B. Performance of ChatGPT on the test of understanding graphs in kinematics. *Phys. Rev. Phys. Educ. Res.* **2024**, *20*, 010109. https://doi.org/10.1103/PhysRevPhysEducRes.20.010109.
39. Kieser, F.; Wulff, P.; Kuhn, J.; Küchemann, S. Educational data augmentation in physics education research using ChatGPT. *Phys. Rev. Phys. Educ. Res.* **2023**, *19*, 020150. https://doi.org/10.1103/PhysRevPhysEducRes.19.020150.
40. Kortemeyer, G. Toward AI grading of student problem solutions in introductory physics: A feasibility study. *Phys. Rev. Phys. Educ. Res.* **2023**, *19*, 020163. https://doi.org/10.1103/PhysRevPhysEducRes.19.020163.
41. Wan, T.; Chen, Z. Exploring generative AI assisted feedback writing for students' written responses to a physics conceptual question with prompt engineering and few-shot learning. *Phys. Rev. Phys. Educ. Res.* **2024**, *20*, 010152. https://doi.org/10.1103/PhysRevPhysEducRes.20.010152.
42. Barakina, E.Y.; Popova, A.V.; Gorokhova, S.S.; Voskovskaya, A.S. Digital technologies and artificial intelligence technologies in education. *Eur. J. Contemp. Educ.* **2021**, *10*, 285–296.
43. Fernández-Herrero, J. Evaluating recent advances in affective intelligent tutoring systems: A scoping review of educational impacts and future prospects. *Educ. Sci.* **2024**, *14*, 839.

44. Mandai, K.; Tan, M.J.H.; Padhi, S.; Pang, K.T. A cross-era discourse on ChatGPT's influence in higher education through the Lens of John Dewey and Benjamin Bloom. *Educ. Sci.* **2024**, *14*, 614.
45. Huesca, G.; Martínez-Treviño, Y.; Molina-Espinosa, J.M.; Sanromán-Calleros, A.R.; Martínez-Román, R.; Cendejas-Castro, E.A.; Bustos, R. Effectiveness of using ChatGPT as a tool to strengthen benefits of the flipped learning strategy. *Educ. Sci.* **2024**, *14*, 660.
46. Dann, C.; O'Neill, S.; Getenet, S.; Chakraborty, S.; Saleh, K.; Yu, K. Improving teaching and learning in higher education through machine learning: Proof of concept' of AI's ability to assess the use of key microskills. *Educ. Sci.* **2024**, *14*, 886.
47. Ali, D.; Fatemi, Y.; Boskabadi, E.; Nikfar, M.; Ugwuoke, J.; Ali, H. ChatGPT in teaching and learning: A systematic review. *Educ. Sci.* **2024**, *14*, 643.
48. Lahme, S.Z.; Klein, P.; Lehtinen, A.; Müller, A.; Pirinen, P.; Rončević, L.; Sušac, A. Physics lab courses under digital transformation: A trinational survey among university lab instructors about the role of new digital technologies and learning objectives. *Phys. Rev. Phys. Educ. Res.* **2023**, *19*, 020159. https://doi.org/10.1103/PhysRevPhysEducRes.19.020159.
49. Kortemeyer, G.; Bauer, W.; Fisher, W. Hybrid teaching: A tale of two populations. *Phys. Rev. Phys. Educ. Res.* **2022**, *18*, 020130. https://doi.org/10.1103/PhysRevPhysEducRes.18.020130.
50. Laumann, D.; Fischer, J.A.; Stürmer-Steinmann, T.K.; Welberg, J.; Weßnigk, S.; Neumann, K. Designing e-learning courses for classroom and distance learning in physics: The role of learning tasks. *Phys. Rev. Phys. Educ. Res.* **2024**, *20*, 010107. https://doi.org/10.1103/PhysRevPhysEducRes.20.010107.
51. Dudar, V.L.; Riznyk, V.V.; Kotsur, V.V.; Pechenizka, S.S.; Kovtun, O.A. Use of modern technologies and digital tools in the context of distance and mixed learning. *Linguist. Cult. Rev.* **2021**, *5*, 733–750.
52. Turgut, Y.E.; Aslan, A. Factors affecting ICT integration in Turkish education: A systematic review. *Educ. Inf. Technol.* **2021**, *26*, 4069–4092.
53. Lacka, E.; Wong, T.; Haddoud, M.Y. Can digital technologies improve students' efficiency? Exploring the role of virtual Learning environment and social media use in higher education. *Comput. Educ.* **2021**, *163*, 104099.
54. Tlili, A.; Zhang, J.; Papamitsiou, Z.; Manske, S.; Huang, R.; Kinshuk; Hoppe, H.U. Towards utilising emerging technologies to address the challenges of using Open Educational Resources: A vision of the future. *Educ. Technol. Res. Dev.* **2021**, *69*, 515–532.
55. Abass, O.A.; Arowolo, O.A.; Igwe, E.N. Towards enhancing service delivery in higher education institutions via knowledge management technologies and blended E-learning. *Int. J. Stud. Educ.* **2021**, *3*, 10–21.
56. Lacka, E.; Wong, T.C. Examining the impact of digital technologies on students' higher education outcomes: The case of the virtual learning environment and social media. *Stud. High. Educ.* **2021**, *46*, 1621–1634.
57. Murod, U.; Suvankulov, B.; Bakiyeva, M.; Nusratova, D. Fundamentals of creation and use of interactive electronic courses on the basis of multimedia technologies. *Ann. Rom. Soc. Cell Biol.* **2021**, *25*, 6860–6865.
58. Syed, A.M.; Ahmad, S.; Alaraifi, A.; Rafi, W. Identification of operational risks impeding the implementation of eLearning in higher education system. *Educ. Inf. Technol.* **2021**, *26*, 655–671.
59. Karimian, Z.; Farrokhi, M.R.; Moghadami, M.; Zarifsanaiey, N.; Mehrabi, M.; Khojasteh, L.; Salehi, N. Medical education and COVID-19 pandemic: A crisis management model towards an evolutionary pathway. *Educ. Inf. Technol.* **2022**, *27*, 3299–3320.
60. Nuere, S.; De Miguel, L. The digital/technological connection with COVID-19: An unprecedented challenge in university teaching. *Technol. Knowl. Learn.* **2021**, *26*, 931–943.
61. Marek, M.W.; Chew, C.S.; Wu, W.-c.V. Teacher experiences in converting classes to distance learning in the COVID-19 pandemic. *Int. J. Distance Educ. Technol.* **2021**, *19*, 89–109.
62. Islam, M.K.; Sarker, M.F.H.; Islam, M.S. Promoting student-centred blended learning in higher education: A model. *E-Learn. Digit. Media* **2022**, *19*, 36–54.
63. Song, H.S.; Kalet, A.L.; Plass, J.L. Interplay of prior knowledge, self-regulation and motivation in complex multimedia learning environments. *J. Comput. Assist. Learn.* **2016**, *32*, 31–50. https://doi.org/10.1111/jcal.12117.
64. Pokhrel, S.; Chhetri, R. A Literature Review on Impact of COVID-19 Pandemic on Teaching and Learning. *High. Educ. Future* **2021**, *8*, 133–141. https://doi.org/10.1177/2347631120983481.
65. Zhang, T.; Taub, M.; Chen, Z. Measuring the Impact of COVID-19 Induced Campus Closure on Student Self-Regulated Learning in Physics Online Learning Modules. In Proceedings of the LAK21: 11th International Learning Analytics and Knowledge Conference, Irvine, CA, USA, 12–16 April 2021; pp. 110–120.
66. Chhetri, C. "I Lost Track of Things": Student Experiences of Remote Learning in the COVID-19 Pandemic. In Proceedings of the 21st Annual Conference on Information Technology Education, Virtual Event, USA, 7–9 October 2020; pp. 314–319.

67. Marshman, E.; DeVore, S.; Singh, C. Holistic framework to help students learn effectively from research-validated self-paced learning tools. *Phys. Rev. Phys. Educ. Res.* **2020**, *16*, 020108. https://doi.org/10.1103/PhysRevPhysEducRes.16.020108.
68. Marshman, E.; DeVore, S.; Singh, C. Challenge of helping introductory physics students transfer their learning by engaging with a self-paced learning tutorial. *Front. ICT* **2018**, *5*, 3. https://doi.org/10.3389/fict.2018.00003.
69. DeVore, S. Using the Tutorial Approach to Improve Physics Learning from Introductory to Graduate Level. Ph.D. Thesis, University of Pittsburgh, Pittsburgh, PA, USA, 2015.
70. DeVore, S.; Marshman, E.; Singh, C. Challenge of engaging all students via self-paced interactive electronic learning tutorials for introductory physics. *Phys. Rev. Phys. Educ. Res.* **2017**, *13*, 010127. https://doi.org/10.1103/PhysRevPhysEducRes.13.010127.
71. Schwartz, D.; Bransford, J.; Sears, D. Efficiency and innovation in transfer. In *Transfer of Learning from a Modern Multidisciplinary Perspective*; Information Age Publishing: Greenwich, CT, USA, 2005; Volume 3, pp. 1–51.
72. Schwartz, D.L.; Bransford, J.D. A time for telling. *Cogn. Instr.* **1998**, *16*, 475–5223.
73. Winne, P. H. Cognition and Metacognition within Self-Regulated Learning. In P. A. Alexander, D. H. Schunk, & J. A. Greene (Eds.), *Handbook of self-regulation of learning and performance (2nd ed.)* (**2018**. 2nd ed., Issue 10449, pp. 36–48). Routledge. https://doi.org/10.4324/9781315697048.ch3
74. Bandura, A. Social cognitive theory of self-regulation. *Organ. Behav. Hum. Decis. Process.* **1991**, *50*, 248–287.
75. Kirkwood, A.; Price, L. Technology-enhanced learning and teaching in higher education: What is 'enhanced' and how do we know? A critical literature review. *Learn. Media Technol.* **2014**, *39*, 6–36. https://doi.org/10.1080/17439884.2013.770404.
76. Choi-Lundberg, D.L.; Butler-Henderson, K.; Harman, K.; Crawford, J. A systematic review of digital innovations in technology-enhanced learning designs in higher education. *Australas. J. Educ. Technol.* **2023**, *39*, 133–162. https://doi.org/10.14742/ajet.7615.
77. Lucas, H.C.; Upperman, J.S.; Robinson, J.R. A systematic review of large language models and their implications in medical education. *Med. Educ.* **2024**, *58*, 1276–1285. https://doi.org/10.1111/medu.15402.
78. van Seters, J.R.; Ossevoort, M.A.; Tramper, J.; Goedhart, M.J. The influence of student characteristics on the use of adaptive e-learning material. *Comput. Educ.* **2012**, *58*, 942–952. https://doi.org/10.1016/j.compedu.2011.11.002.
79. Bransford, J.D.; Schwartz, D.L. Rethinking Transfer: A Simple Proposal with Multiple Implications. *Rev. Res. Educ.* **1999**, *24*, 61. https://doi.org/10.2307/1167267.
80. Taub, M.; Banzon, A.M.; Zhang, T.; Chen, Z. Tracking changes in students' online self-regulated learning behaviors and achievement goals using trace clustering and process mining. *Front. Psychol.* **2022**, *13*, 813514. https://doi.org/10.3389/fpsyg.2022.813514.
81. Chen, Z.; Xu, M.; Garrido, G.; Guthrie, M.W. Relationship between students' online learning behavior and course performance: What contextual information matters? *Phys. Rev. Phys. Educ. Res.* **2020**, *16*, 010138. https://doi.org/10.1103/PhysRevPhysEducRes.16.010138.
82. Whitcomb, K.M.; Guthrie, M.W.; Singh, C.; Chen, Z. Improving accuracy in measuring the impact of online instruction on students' ability to transfer physics problem-solving skills. *Phys. Rev. Phys. Educ. Res.* **2021**, *17*, 010112. https://doi.org/10.1103/PhysRevPhysEducRes.17.010112.
83. Doucette, D.; Cwik, S.; Singh, C. "Everyone is new to this": Student reflections on different aspects of online learning. *Am. J. Phys.* **2021**, *89*, 1042–1047.
84. Available online: https://www.mooc.org/ (accessed on 1 September 2024)
85. Bego, C.R.; Lyle, K.B.; Ralston, P.A.S.; Immekus, J.C.; Chastain, R.J.; Haynes, L.D.; Hoyt, L.K.; Pigg, R.M.; Rabin, S.D.; Scobee, M.W.; et al. Single-paper meta-analyses of the effects of spaced retrieval practice in nine introductory STEM courses: Is the glass half full or half empty? *Int. J. STEM Educ.* **2024**, *11*, 9. https://doi.org/10.1186/s40594-024-00468-5.
86. Felker, Z.; Chen, Z. Reducing procrastination on introductory physics online homework for college students using a planning prompt intervention. *Phys. Rev. Phys. Educ. Res.* **2023**, *19*, 010123. https://doi.org/10.1103/PhysRevPhysEducRes.19.010123.
87. Yan, L.; Greiff, S.; Teuber, Z.; Gašević, D. Promises and challenges of generative artificial intelligence for human learning. *Nat. Hum. Behav.* **2024**, *8*, 1839–1850. https://doi.org/10.1038/s41562-024-02004-5.
88. Kasneci, E.; Seßler, K.; Küchemann, S.; Bannert, M.; Dementieva, D.; Fischer, F.; Gasser, U.; Groh, G.; Günnemann, S.; Hüllermeier, E.; et al. ChatGPT for Good? On Opportunities and Challenges of Large Language Models for Education. *Learn. Individ. Differ.* **2023**, *103*, 102274. https://doi.org/10.1016/j.lindif.2023.102274.
89. Küchemann, S.; Steinert, S.; Kuhn, J.; Avila, K.; Ruzika, S. Large language models—Valuable tools that require a sensitive integration into teaching and learning physics. *Phys. Teach.* **2024**, *62*, 400–402. https://doi.org/10.1119/5.0212374.
90. Mirzadeh, I.; Alizadeh, K.; Shahrokhi, H.; Tuzel, O.; Bengio, S.; Farajtabar, M. GSM-Symbolic: Understanding the Limitations of Mathematical Reasoning in Large Language Models. *arXiv* **2024**, arXiv:2410.05229.

91. Yan, Z.; Zhang, R.; Jia, F. Exploring the Potential of Large Language Models as a Grading Tool for Conceptual Short-Answer Questions in Introductory Physics. In Proceedings of the 2024 9th International Conference on Distance Education and Learning, Guangzhou, China, 14–17 June 2024; pp. 308–314.
92. Sonkar, S.; Liu, N.; Mallick, B.D.; Baraniuk, G.R. Marking: Visual Grading with Highlighting Errors and Annotating Missing Bits. *arXiv* **2024**, arXiv:2404.14301.
93. Giray, L. Prompt Engineering with ChatGPT: A Guide for Academic Writers. *Ann. Biomed. Eng.* **2023**, *51*, 2629–2633. https://doi.org/10.1007/s10439-023-03272-4.
94. Latif, E.; Zhai, X. Fine-tuning ChatGPT for automatic scoring. *Comput. Educ. Artif. Intell.* **2024**, *6*, 100210. https://doi.org/10.1016/j.caeai.2024.100210.
95. Carpenter, D.; Min, W.; Lee, S.; Ozogul, G.; Zheng, X.; Lester, J. Assessing Student Explanations with Large Language Models Using Fine-Tuning and Few-Shot Learning. In Proceedings of the 19th Workshop on Innovative Use of NLP for Building Educational Applications, Mexico City, Mexico, 20–21 June 2024; pp. 403–413.
96. Lewis, P.; Perez, E.; Piktus, A.; Petroni, F.; Karpukhin, V.; Goyal, N.; Küttler, H.; Lewis, M.; Yih, W.; Rocktäschel, T.; et al. Retrieval-Augmented Generation for Knowledge-Intensive NLP Tasks. *Adv. Neural Inf. Process. Syst.* **2020**, *33*, 9459–9474.
97. Available online: https://www.khanacademy.org/ (accessed on 1 September 2024).
98. Available online: https://chrisyandata.medium.com/understanding-temperature-setting-in-generative-ai-models-be65489b82fd (accessed on 1 September 2024).
99. Palincsar, A.S.; Brown, A.L. Reciprocal teaching of comprehension-fostering and comprehension-monitoring activities. *Cogn. Instr.* **1984**, *1*, 117–175. https://doi.org/10.1207/s1532690xci0102_1.
100. Hsu, L.; Brewe, E.; Foster, T.M.; Harper, K.A. Resource Letter RPS-1: Research in problem solving. *Am. J. Phys.* **2004**, *72*, 1147–1156. https://doi.org/10.1119/1.1763175.
101. Reif, F.; Scott, L.A. Teaching scientific thinking skills: Students and computers coaching each other. *Am. J. Phys.* **1999**, *67*, 819–831. https://doi.org/10.1119/1.19130.
102. Eke, D.O. ChatGPT and the rise of generative AI: Threat to academic integrity? *J. Responsible Technol.* **2023**, *13*, 100060. https://doi.org/10.1016/j.jrt.2023.100060.
103. Rozado, D. The Political Biases of ChatGPT. *Soc. Sci.* **2023**, *12*, 148.
104. Li, L.; Bamman, D.; Akoury, N.; Brahman, F.; Chaturvedi, S.; Clark, E.; Iyyer, M.; Martin, L.J. Gender and Representation Bias in GPT-3 Generated Stories. In Proceedings of the Third Workshop on Narrative Understanding, Association for Computational Linguistics, Virtual, 11 June 2021. https://doi.org/10.18653/v1/2021.nuse-1.5.
105. Zhang, M.; Press, O.; Merrill, W.; Liu, A.; Smith, N.A.; Allen, P.G. How Language Model Hallucinations Can Snowball. *arXiv* **2023**, arXiv:2305.13534.
106. Pădurean, V.-A.; Denny, P.; Singla, A. BugSpotter: Automated Generation of Code Debugging Exercises. *arXiv* **2024**, arXiv:2411.14303.
107. Tan, S.C.; Chen, W.; Chua, B.L. Leveraging generative artificial intelligence based on large language models for collaborative learning. *Learn. Res. Pract.* **2023**, *9*, 125–134.
108. Barana, A.; Marchisio, M.; Roman, F. Fostering problem solving and critical thinking in mathematics through generative artificial intelligence. In Proceedings of the 20th international conference on Cognition and Exploratory Learning in the Digital Age (CELDA 2023), Isola di Madeira, Portugal, 21–23 October 2023.
109. Moongela, H.; Matthee, M.; Turpin, M.; van der Merwe, A. The Effect of Generative Artificial Intelligence on Cognitive Thinking Skills in Higher Education Institutions: A Systematic Literature Review. In *Artificial Intelligence Research*; Springer: Cham, Switzerland, 2024; pp. 355–371.